\documentclass[%
 reprint,
 superscriptaddress,
 amsmath,amssymb,
 aps,
]{revtex4-2}

\usepackage{amsmath}
\usepackage{printlen}
\usepackage{graphicx}
\usepackage{dcolumn}
\usepackage{bm}
\usepackage{subfigure}
\usepackage{xcolor} 
\usepackage{amssymb} 
\usepackage{verbatim} 
\usepackage{lmodern}
\usepackage{float} 
\usepackage{hyperref}
\usepackage[normalem]{ulem}

\newcommand{\vv}{\vec}

\begin{document}

\preprint{APS/123-QED}

\title{Yielding under compression and the polyamorphic transition in silicon}

\author{Jan Grießer}
\affiliation{Department of Microsystems Engineering, University of Freiburg, Georges-K\"ohler-Allee 103, 79110 Freiburg, Germany}
\author{Gianpietro Moras}
\affiliation{Fraunhofer IWM, MicroTribology Center $\mu$TC, W\"ohlerstra{\ss}e 11, 79108 Freiburg, Germany}
\author{Lars Pastewka}
\email[Corresponding author: ]{lars.pastewka@imtek.uni-freiburg.de}
\affiliation{Department of Microsystems Engineering, University of Freiburg, Georges-K\"ohler-Allee 103, 79110 Freiburg, Germany}
\affiliation{Cluster of Excellence livMatS, Freiburg Center for Interactive Materials and Bioinspired Technologies, University of Freiburg, 79110 Freiburg, Germany}

\date{\today}

\begin{abstract}
We investigate the behavior of amorphous silicon under hydrostatic compression using molecular simulations.
During compression, amorphous silicon undergoes a discontinuous nonequilibrium transition from a low-density to a high-density structure at a pressure of around $13$-$16$~GPa.
Ensemble-averaged density and elastic constants change discontinuously across the transition.
Densification of individual glassy samples occurs through a series of discrete plastic events, each of which is accompanied by a vanishing shear modulus.
This is the signature of a series of elastic instabilities, similar to shear transformation zones observed during shear yielding of glasses.
We compare the structure obtained during compression with a near-equilibrium form of amorphous silicon obtained by quenching a melt at constant pressure.
This gives structures identical to nonequilibrium compression at low and high pressure, but the transition between them occurs gradually rather than discontinuously.
Our observations indicate that the polyamorphic transition is of a nonequilibrium nature, and it has the characteristics of a yield transition that occurs under compression instead of shear.

\end{abstract}

\maketitle

\section{Introduction}
Materials can exist in structurally distinct forms in their crystalline state, a property which is called \emph{polymorphism}~\cite{callister_2007}.
Which form is actually present depends on the external conditions temperature and pressure and on the way the material is formed and processed afterwards.
One form can transform into another, for example when increasing the pressure above some critical value.
In crystals, pressure-induced phase transitions are commonly between equilibrium states of the system~\cite{stanley_1971,kubo_statistical_physics1, schwabl_2006}.
Amorphous solids, however, are intrinsically out-of-equilibrium, yet pressure-induced transformations between distinct amorphous forms are possible~\cite{poole_polymorphism_1997,brazhkin_AAT_2003,wilding_structural_and_polymorpishm_2006,mcmillan_polyamorphism_2007,machon_PIA_AAT_2014,tanaka_polyamorphism_2020}.

\begin{figure*}[h!t!]
	\includegraphics[]{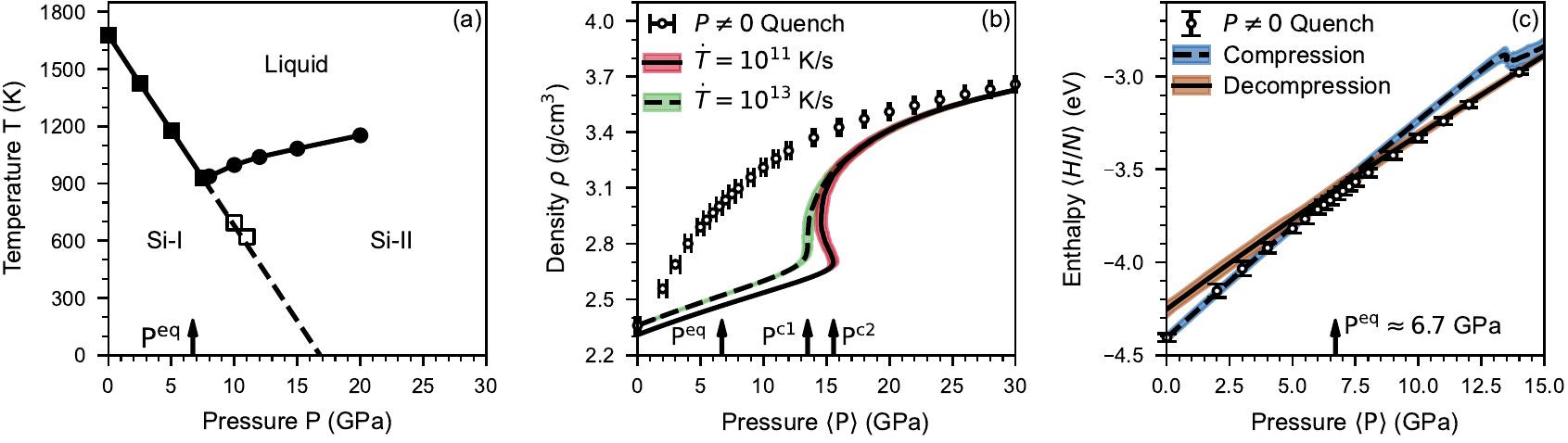}
	\caption{
     (a) Pressure-temperature phase diagram of silicon from phase-coexistence calculations (Data from Ref.~\cite{moras_disappearance_polyamorphic_2018}.)
     Si-I is the diamond cubic and Si-II the $\beta$-tin phase of silicon.
     (b) Density of amorphous silicon as a function of hydrostatic pressure.
     The dashed and solid line are glasses hydrostatically compressed in AQC simulations and initially prepared at vanishing pressure.
     Dots represent glasses quenched from the liquid while maintaining the constant non-zero hydrostatic pressure. 
     Lines and dots indicate the ensemble mean while shaded areas and error bars mark the standard deviation.
     The vertical arrow $P^{\text{eq}}$ marks the critical pressure for the expected equilibrium transition (see panel c for its definition), $P^{\text{c1}} = 13.5\,$GPa and $P^{\text{c2}} = 15.6\,$GPa mark the critical pressures for the nonequilibrium transition.
     (c) Enthalpy during AQC simulations of compression and subsequent decompression of an ensemble of glasses prepared with quench rate of $\dot{T}=10^{13}\,$K/s. 
     The vertical arrow $P^{\text{eq}}\approx 6.7\,$GPa marks the intersection of the enthalpy curves during compression and decompression.}
	\label{fig: hydrostatic_compression}
\end{figure*}

Materials with multiple amorphous modifications are called \emph{polyamorphic}~\cite{poole_polymorphism_1997,brazhkin_AAT_2003,wilding_structural_and_polymorpishm_2006,mcmillan_polyamorphism_2007,tanaka_polyamorphism_2020}.
One prominent example of a material featuring such a polyamorphic phase transition is silicon~\cite{deb_pressure_2001,durandurdu_abInitio_amorphous_2001,sastry_liquid_liquid_silicon_2003,morishita_high_density_silicon_2004,mcmillan_density_driven_transition_2005,garg_memory_effect_LDA_2011,daisenberger_polyamorphic_2011,moras_disappearance_polyamorphic_2018,deringer_origins_polyamorphous_2021}.
Silicon crystallizes into the diamond cubic phase under ambient conditions and transforms into the $\beta$-Sn phase upon hydrostatic compression~\cite{haberl_inisght_defectrole_aSi_2013}.
An indication for a polyamorphic transition in the corresponding amorphous form is that the melting temperature $T^{\text{m}}$ in silicon decreases with increasing pressure, $dT^{\text{m}}/dP < 0$ (see Fig.~\ref{fig: hydrostatic_compression}a), meaning that the liquid phase is denser than the low-temperature diamond cubic crystal.
A maximum in the melting curve, which for silicon is expected to occur at negative pressure~\cite{mcmillan_polyamorphic_transitions_2004,wilding_structural_and_polymorpishm_2006,vasisht_silicon_LLT_2011}, is typically explained by a ``two-state'' model~\cite{rapoport_1967}.
This model assumes that low-density and high-density phases coexist in the liquid state, with the proportion of high-density domains increasing with pressure, thus leading to a pressure-driven transition from a low-density to a high-density liquid phase~\cite{sastry_liquid_liquid_silicon_2003,mcmillan_polyamorphic_transitions_2004,wilding_structural_and_polymorpishm_2006,tanaka_polyamorphism_2020}.
A transition between two amorphous modifications is expected to reflect this liquid-liquid phase transition.

Experimental studies revealed that amorphous silicon (aSi) can indeed exist in two distinct structures.
Upon compression, aSi transforms from a low-density structure (LDA) to a high-density structure (HDA) at a critical pressure between $P^\text{c} \approx 12\,$GPa and $16\,$GPa~\cite{deb_pressure_2001,mcmillan_density_driven_transition_2005,daisenberger_high_2007,pandey_pressure_induced_crystallization_2011}.
These experiments were supported by early numerical simulations, which showed a transition from an initially four-fold coordinated LDA structure to an HDA structure with five-fold coordinated atoms~\cite{morishita_high_density_silicon_2004}, or to a very high density amorphous (VHDA) structure with even higher coordination~\cite{durandurdu_abInitio_amorphous_2001}.
This structural transformation of aSi has been numerically investigated by multiple authors since, always showing a coexistence of LDA and HDA regions that sequentially transform to HDA (or VHDA) during compression~\cite{durandurdu_abInitio_amorphous_2001,morishita_high_density_silicon_2004,moras_disappearance_polyamorphic_2018,deringer_origins_polyamorphous_2021}.

There are multiple well-documented mechanisms that lead to structural transitions between polymorph.
Besides equilibrium phase transitions, crystalline materials have ultimate stability limits where the crystalline lattice collapses as a consequence of an elastic or dynamic instability~\cite{wallace_1972,hill_elasticity_stability_1975,wang_stability_1995,griesser_elastic_constants_2023}.
The result of such an instability is that all atoms in a perfect and defect-free crystal rearrange at once, forming a new stable crystal.
Plasticity during shear loading of amorphous structures is associated with similar, yet localized mechanical instabilities~\cite{lacks_localized_instabilities_1998,malandro_1999,maloney_breakdown_2004,lemaitre_sum_rules_2006,Cubuk2017-vo,richard_plasticity_2020}.
The rearranging regions are called shear-transformation zones and their occurrence can be predicted through the identification of ``soft spots''~\cite{falk_dynamics_1998,ding_soft_spots_2014,Cubuk2017-vo,richard_plasticity_2020}.
In aSi, these soft spots have been identified as local regions of HDA character~\cite{demkowicz_plastic_carriers1_2004,demkowicz_plastic_carries2_2005}.
This raises the question whether plasticity in amorphous solids and the polyamorphic transition have different microscopic signatures, or conversely, whether this transition is nothing else than some sort of yield point.

In the theory of equilibrium phase transitions, the distinction between first and second order is made on the behavior of susceptibilities across the transition point~\cite{huang_statistical_physics,schwabl_statistical_2006}.
A partial answer to the question on the nature of the polyamorphic transition may therefore rest in how the (linear) mechanical response of aSi behaves across the transition.
We here use atomic-scale computer simulations of periodic representative amorphous volume elements, to study the macroscopic (elastic) and microscopic (shear transformation) response of aSi across the polyamorphic transition.
To drive the transition, we compress (and decompress) these volume elements affinely in small increments, and search for the next local minimum after the deformation step.
We call this procedure athermal quasistatic \emph{compression} (AQC), in analogy to the athermal quasistatic \emph{shear} procedure used in the amorphous plasticity literature~\cite{maeda_athermal_1981,maloney_athermal_2006}.
It mimicks experiments where pressure is induced either through a pressure medium in a loading cell or by contacting with an indenter~\cite{deb_pressure_2001,mcmillan_density_driven_transition_2005,daisenberger_high_2007,pandey_pressure_induced_crystallization_2011,garg_memory_effect_LDA_2011,haberl_inisght_defectrole_aSi_2013}.

Our results indicate, that the collapse of the aSi structure has signatures similar to the yield point during shear loading.
The transition can be viewed as a sequential series of spatially localized compression transformation zones or plastic events.
To emphasize the nonequilibrium character of the transition, we also probe the properties of a form of aSi that is compressed in the liquid phase and then quenched at constant pressure to become amorphous.
The resulting amorphous samples show a gradual, rather than a discontinuous, variation of thermodynamic properties (density, elastic constants) as a function of pressure.

\section{Methods}
We employ molecular dynamics and molecular statics simulations to investigate the behavior of aSi under hydrostatic compression.
We first melt a Si-I (diamond cubic) crystal with $N=4096$ atoms in the isothermal-isobaric (NPT) ensemble at $T=3000\,$K and zero hydrostatic pressure. 
The liquid is cooled down to $T=2000\,$K at a constant rate of $\dot{T}=10^{13}\,$K/s and equilibrated for further $20\,$ps.

Our final aSi configurations are prepared from this silicon melt in two distinct ways.
First, we produce periodic representative volume elements of aSi at zero hydrostatic pressure.
We draw new random velocities from a Gaussian distribution at $T=2000\,$K and equilibrate the structures for $30\,$ps.
This equilibration time is sufficient for the atoms to loose memory of their initial state i.e. for the mean-squared displacement to be in the diffusive regime.
After equilibration, we quench the liquid samples from $T=2000\,$K to $T=1\,$K at quench rates of $\dot{T}=10^{13}\,$K/s and $\dot{T}=10^{11}\,$K/s.
To drive the polyamorphic transition, the final glass is then compressed hydrostatically by affinely remapping all atomic positions with a prescribed density increment $\Delta \rho$~\cite{maeda_athermal_1981,maloney_athermal_2006}.
We optimize the glass structure after each increment, a procedure we refer to as AQC in the following.
Second, we equilibrate liquid silicon at $T=2000\,$K and a set of constant hydrostatic pressures $P \neq 0\,$GPa.
The liquid is quenched to $T=1\,$K using a quench rate of $\dot{T}=10^{13}\,$K/s while maintaining the constant hydrostatic pressure.
We generated ensembles of $500$ independent configurations for the glasses quenched at vanishing pressure and $100$ configurations for each quench at constant pressure.
In all simulations, we used the interatomic potential by Kumagai et al.~\cite{kumagai_potential_2007}.
Previous publications showed that this interatomic potential correctly reproduces the polyamorphic transition in aSi~\cite{moras_disappearance_polyamorphic_2018} and delivers results on shear-induced silicon amorphization that are comparable to those obtained with more complex machine-learning potentials~\cite{reichenbach_amorphization_2021, bartok2018machine}.
All dynamic simulations employed a timestep $\Delta t = 1.0\,$fs, a Nose-Hoover chain thermostat and Parinello-Rahman barostat~\cite{parrinello_barostat_1981,martyna_constant_pressure_1994,shinoda_2004}.
The relaxation constant of the thermostat and the barostat were chosen as $1\,$ps and $10\,$ps, respectively. 
Energy minimization in static simulations was performed using a conjugate-gradient minimizer with a force tolerance of $10^{-6}\,$eV/$\text{\AA}$.

\section{Results}
\subsection{Phenomenology}
We begin by investigating how amorphous silicon prepared at zero pressure behaves under hydrostatic compression.
As shown in Fig.~\ref{fig: hydrostatic_compression}b, the density initially increases linearly with pressure.
The material densifies at constant critical pressure $P^{\text{c}}$, which manifests as an abrupt increase in the density-pressure diagram.
This sudden compression is the signature of the pressure-driven transition from an LDA to an HDA structure in aSi.
The critical pressure for this transition is at $P^{\text{c1}} \approx 13.5\,$GPa and $P^{\text{c2}} \approx 15.6\,$GPa for glasses prepared with quench rates of $10^{13}\,$K/s and $10^{11}\,$K/s, respectively.
As the system is further compressed, the mean density of glasses prepared with different quench rates converges to the same value.
The glasses prepared at a constant non-zero pressure show a smooth dependence on the hydrostatic pressure and therefore a continuous transition between the two phases.
These results are consistent with prior work~\cite{moras_disappearance_polyamorphic_2018,deringer_origins_polyamorphous_2021}.

Additional insight can be obtained by comparing the enthalpy $H$ of the two phases~\cite{kubo_statistical_physics1,lacks_silica_2000,durandurdu_abInitio_amorphous_2001},
$H = U + PV$,
where $U$ is the potential energy, $P$ is the pressure and $V$ is the volume.
In Fig.~\ref{fig: hydrostatic_compression}c, we show the enthalpy during compression and subsequent decompression.
The two curves intersect at $P^{\text{eq}}\approx 6.7\,$GPa, which is the pressure at which an ``equilibrium'' phase transition could occur.
Glasses prepared at constant pressure follow the amorphous structure with the lowest enthalpy.
We conclude that below $6.7\,$GPa the LDA structure is thermodynamically preferred, while above $6.7\,$GPa the HDA structure is more favorable.
The critical pressures $P^{\text{c}}$ obtained from compression are more than twice this ``equilibrium'' pressure $P^{\text{eq}}$.

\begin{figure}[t!]
\includegraphics[]{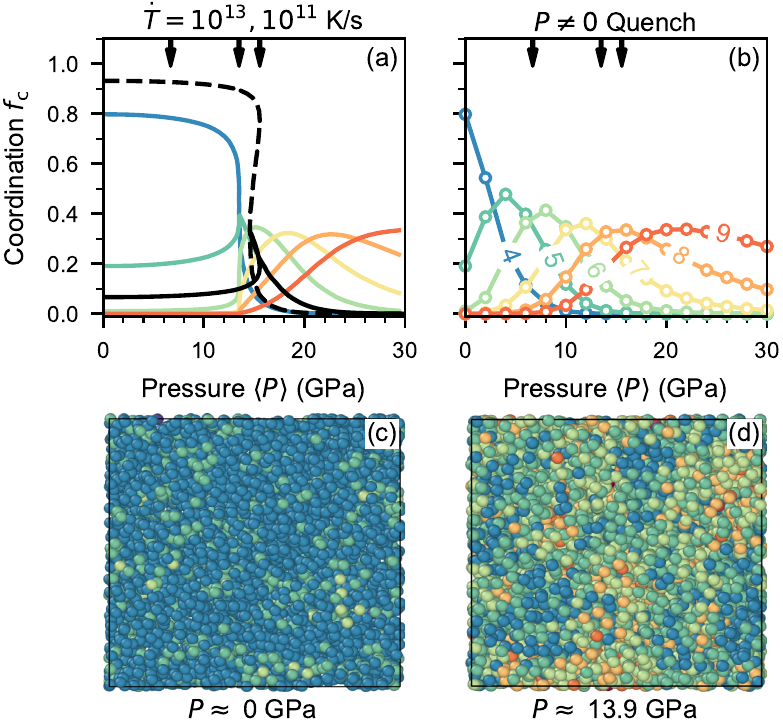}
    \caption{\label{fig: coordination_number}
    (a,b) Probability $f_c$ of finding an atom with coordination $c$ as a function of hydrostatic pressure for glasses prepared at zero pressure (a) and at constant nonzero pressure (b) from ensembles of simulation runs.
    Colored lines mark samples prepared initially with a quench rate of $\dot{T}=10^{13}\,$K/s.
    The black dashed line shows $f_4$ and the black solid line $f_5$ for glasses prepared with a quench rate of $\dot{T}=10^{11}\,$K/s.
    We only show $f_4$ and $f_5$ for the latter since atoms with higher coordination behave similar for compression larger than the critical pressure.
    The spatial cutoff for the computation of the coordination number is chosen as $r_\text{c}=2.925\,\text{\AA}$ as obtained from the first local minimum in the radial distribution function at zero pressure. 
    The vertical arrows indicate $P^{\text{eq}}$, $P^{\text{c1}}$ and $P^{\text{c2}}$ as in Fig.~\ref{fig: hydrostatic_compression}.
    (c,d) Visualization of the coordination number $c$ for an exemplary atomic configuration quenched with $\dot{T}=10^{13}\,$K/s in the LDA (c) and the HDA structure (d).
    Color coding is the same as in panels (a) and (b).}
\end{figure}

We now analyze how the atomic structure changes across this transition.
Figures~\ref{fig: coordination_number}a and b show the probability $f_c$ of finding an atom with coordination $c$.
For glasses prepared at a quench rate of $10^{13}\,$K/s we show $c\in \{4,...,9\}$ while we only show $c \in \{4, 5\}$ for the slower quench rate of $10^{11}\,$K/s.
During hydrostatic compression, before $P^{\text{c}}$ is reached, we observe only small changes in the mean coordination $\sum_c c f_c$.
In this regime, slowly quenched glasses have a larger fraction $f_4$ of four-fold coordinated atoms than samples prepared with a fast quench rate.
Upon reaching the critical pressure, $f_4$ drops and an atomic configuration with mainly five-fold coordinated atoms (large $f_5$) emerges.
These five-fold coordinated configurations are transient and vanish rapidly as atoms with even higher coordination numbers 6, 7, 8 and 9 emerge continuously for increasing hydrostatic pressure.
The snapshots of compressed configurations in Fig.~\ref{fig: coordination_number}c (LDA) and d (HDA) also clearly show this increase in average coordination across the LDA-HDA transition.
Note that some authors denote structures with high coordination numbers as VHDA~\cite{durandurdu_abInitio_amorphous_2001,morishita_high_density_silicon_2004,garcez_VHDA_2014,deringer_origins_polyamorphous_2021}, but given the gradual transition towards highly coordinated structures we refer to the high pressure structures as HDA.

For glasses quenched at a constant pressure, the coordination number changes continuously with hydrostatic pressure.
Thereby, low-coordinated atoms are continuously replaced by atoms with higher coordination as the pressure increases.
At the pressure $P^{\text{eq}} = 6.7\,$GPa, the configurations consist mainly of five- and six-fold coordinated atoms. 

\subsection{Elastic properties}

The abrupt change in coordination during compression, as opposed to a continuous pressure dependency, strengthens the assumption of a nonequilibrium phase transition during compression.
To further substantiate this hypothesis, we investigate the elastic properties across the polyamorphic transition.
It is well known that in crystalline materials a mechanically driven transition occurs when an eigenvalue of the elastic tensor $C_{\alpha\beta\mu\nu}$ disappears~\cite{born_dynamical_theory_1955,wallace_stability_crystal_1965,hill_elasticity_stability_1975}.
In the limit of zero temperature and finite stress, the tensor of elastic moduli $C_{\alpha\beta\mu\nu}$ (also known as the Birch coefficients~\cite{birch_finite_1947,birch_elasticity_1952,wang_stability_1995,griesser_elastic_constants_2023}) is composed of three contributions: the affine or Born moduli $C^{\text{B}}_{\alpha\beta\mu\nu}$, the non-affine moduli $C^{\text{NA}}_{\alpha\beta\mu\nu}$ and a stress dependent contribution $C^{\text{S}}_{\alpha\beta\mu\nu}$.
The full elastic tensor is computed as \cite{barron_second_order_elastic_constants_1965,wallace_thermoelasticity_stressed_1967,lutsko_generalized_expressions_elastic_1989,griesser_elastic_constants_2023}
\begin{equation}
    C_{\alpha\beta\mu\nu}
    =
    C^{\text{B}}_{\alpha\beta\mu\nu}
    +
    C^{\text{NA}}_{\alpha\beta\mu\nu}
    +
    C^{\text{S}}_{\alpha\beta\mu\nu}
    \label{eq: elastic_contributions}
\end{equation}
with
\begin{align}
    \begin{split}
      C^{\text{B}}_{\alpha\beta\mu\nu} 
      &=
      \frac{1}{V}
      \frac{\partial^2 U(r_{ij})}{\partial \eta_{\alpha \beta} \partial \eta_{\mu \nu}}  
    \end{split}
    \\
    \begin{split}
      C^{\text{NA}}_{\alpha\beta\mu\nu}
      &=
      -\frac{1}{V}
      \frac{\partial^2 U(r_{ij})}{\partial r_{l\gamma}\partial \eta_{\alpha \beta}}
      \left( 
      \mathcal{H}^{-1}
      \right)_{l\gamma, m\kappa}
      \frac{\partial^2 U(r_{ij})}{\partial r_{m\kappa}\partial \eta_{\mu \nu}}
    \end{split}
    \\
    \begin{split}
      C^{\text{S}}_{\alpha\beta\mu\nu}
      &=
      \frac{1}{2}
      (
      \sigma_{\alpha\mu} \delta_{\beta\nu}
      +
      \sigma_{\alpha\nu} \delta_{\beta\mu}
      +
      \sigma_{\beta\mu} \delta_{\alpha\nu}\\
      &\qquad 
      +
      \sigma_{\beta\nu} \delta_{\alpha\mu}
      -
      2\sigma_{\alpha\beta} \delta_{\mu\nu}
      ),
      \label{eq: stress_dependent_C}
    \end{split}
\end{align}
where $U(r_{ij})$ is the energy function, $V$ is the current volume of the simulation cell, $r_{i\gamma}$ is the position of atom $i$ in direction $\gamma$, $\eta_{\alpha \beta}$ is the Green-Lagrange strain tensor, $\sigma_{\alpha \beta}$ is the Cauchy stress, $\delta_{\alpha \beta}$ is the Kronecker delta and $\mathcal{H} = \partial^2 U(r_{ij})/\partial r_{l\phi} \partial r_{m\psi}$ is the Hessian matrix.
More details on this decomposition at finite stress and its analytical computation for many-body potentials can be found in Ref.~\cite{griesser_elastic_constants_2023}.

Since amorphous solids show isotropic material behavior for sufficiently large system sizes and we consider a hydrostatic stress, the elastic tensor $C_{\alpha\beta\mu\nu}$ reduces to two independent elastic constants~\cite{barron_second_order_elastic_constants_1965,tanguy_continuum_limit_2002,tsamados_local_elasticiy_2009,mizuno_elastic_2013}.
We report the ensemble-averaged bulk modulus $K$ and shear modulus $\mu$, since they are the eigenvalues of the elastic tensor.
They determine the limit of elastic stability that is reached when either modulus vanishes~\cite{wallace_1972,hill_elasticity_stability_1975}.
Only the shear modulus can vanish under hydrostatic compression.

\begin{figure}[t!]
    \includegraphics[]{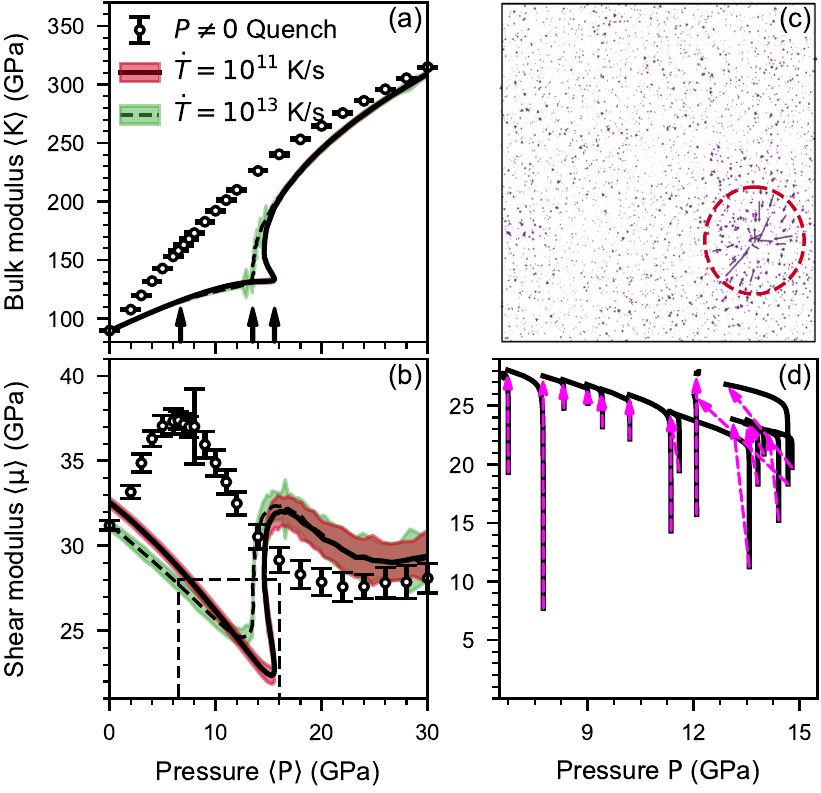}
    \caption{
     Pressure dependent ensemble-averaged bulk modulus $\langle K \rangle$ (a) and shear modulus $\langle \mu \rangle$ (b). 
     Exemplary non-affine displacement field resulting from one elastic instability (c).
     Note that the displacement vectors are scaled with a constant value $x$ for better visibility and the red circle is a guide to the eye and marks the location of largest displacement. 
     (d) Pressure-dependent shear modulus $\mu$ of the black dashed box in panel (b) for one single configuration prepared using $\dot{T}=10^{11}\,$K/s.
     }
    \label{fig: KG_coefficients}
\end{figure}

Figure~\ref{fig: KG_coefficients} shows the bulk modulus $K$ and shear modulus $\mu$ for the full range of pressures.
The bulk modulus of the glasses increases under compression for the full pressure range, independent of preparation.
For aSi prepared at vanishing pressure, the bulk modulus initially increases linearly up to $P \approx 8\,$GPa and starts to become independent of pressure up to $P^\text{c}$.
During the transformation (at $P^\text{c}$), the bulk modulus increases discontinuously and subsequently approaches again a linear pressure dependency for higher compression.
For glasses prepared at constant, nonzero pressure we observe a continuous, but sublinear, increase of the bulk modulus with increasing pressure.
The pressure dependency of the bulk modulus resembles the pressure dependency of the density in Fig.~\ref{fig: hydrostatic_compression}b.
The difference between the quenched and AQC glasses is purely due to their difference in density.
The bulk modulus data collapses when plotted versus density rather than pressure.

The shear modulus $\mu$ shows more interesting features.
During AQC, the shear modulus $\mu$ decreases to a value of about $24\,$GPa at the critical pressure.
Once the $P^{\text{c}}$ is reached, the modulus jumps to a value of roughly $32\,$GPa, almost identical to the modulus for the stress-free quenched structure.
While the initial quench rate leads to a different modulus at $P<P^\text{c}$, the systems appear to loose memory of the initial state beyond the polyamorphic transition.
At even larger compression, the shear modulus starts to soften again.
The shear modulus is independent of preparation protocol above the critical pressure of the polyamorphic transition.

\begin{figure}[t!]
    \includegraphics[]{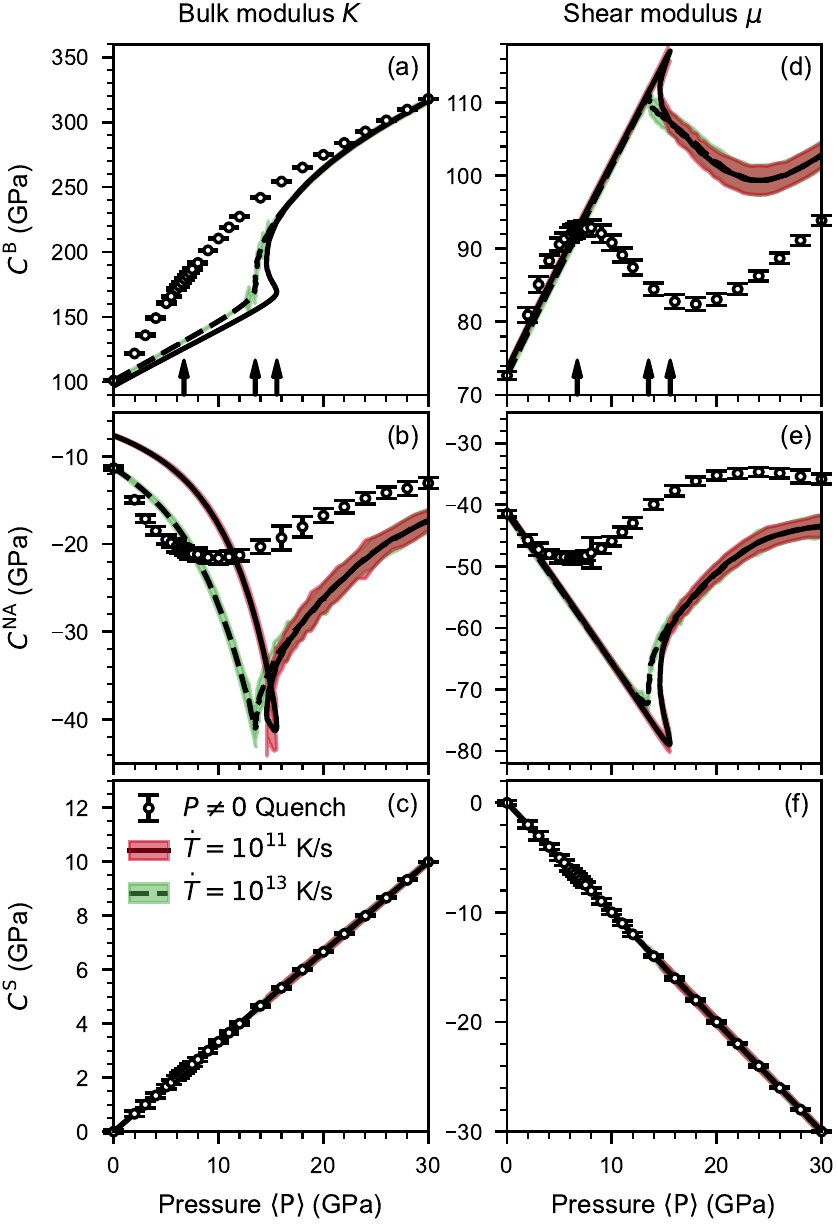}
    \caption{\label{fig:contribution_Kmu}
    Born $C^{\text{B}}$, stress $C^{\text{S}}$ and non-affine $C^{\text{NA}}$ contribution of the bulk modulus $\langle K \rangle$ (a)-(c) and the shear modulus $\langle \mu \rangle$ (d)-(f).}
\end{figure}

For glasses quenched at constant pressure, the shear modulus also has nonmonotonic functional dependency on $P$.
The maximum shear modulus is at $P \approx 6.8$~GPa, which coincides with the equilibrium critical pressure $P^{\text{eq}}$.
As the pressure increases above $P^{\text{eq}}$ and $P^{\text{c}}$, the shear modulus decreases and plateaus at $P\gtrsim 20\,$GPa.
In Figure~\ref{fig:contribution_Kmu} we show the individual contributions from Eq.~\eqref{eq: elastic_contributions} to the bulk modulus and the shear modulus.
For AQC aSi, the pressure dependency of the Born contribution to the bulk modulus is qualitatively similar to what is observed for the full bulk modulus in Fig.~\ref{fig: KG_coefficients}a.
While the Born contribution hardens the elastic response, the non-affine contribution softens it.
With increasing compression, the magnitude of the non-affine contribution increases initially and reaches a local maximum at the critical pressure.
At the critical pressure, the ratio between the Born and the non-affine contribution has a local minimum of $C^\text{B}/ \vert C^{\text{NA}} \vert \approx 4$.
This minimum explains the flattening of the bulk modulus observed in Fig.~\ref{fig: KG_coefficients}a.
Above the transformation, the non-affine contribution decreases its magnitude with pressure.
The aSi samples prepared at a constant pressure show a pressure-dependency similar to the AQC glasses for both the Born and the non-affine contribution. 
However, their pressure dependency is continuous and no abrupt change of either property is observed.  
As expected from Eq.~\eqref{eq: stress_dependent_C}, the stress contribution increases linearly and is independent of the preparation protocol.
The stress contribution additionally stiffens the material, but its absolute value is small compared to the Born contribution. 

The Born contribution to the shear modulus increases linearly under hydrostatic compression for the aSi samples prepared at vanishing pressure.
The behavior of the non-affine contribution is the reserve of the Born contribution, i.e. the magnitude of the non-affine contribution decreases linearly.
At the critical pressure for the transition, the two contributions have a local maximum, or respectively minimum, with a ratio $C^\text{B}/ \vert C^{\text{NA}} \vert \approx 1.4$: The non-affine contribution almost compensation the Born term. 
At the transformation, the pressure dependency changes abruptly.
For larger compression, the affine contribution deceases and the non-affine contribution increases.
The stress-dependent part of the shear modulus softens the elastic response in addition to the non-affine contribution.
This ultimately explains the observed softening of the shear modulus in Fig.~\ref{fig: KG_coefficients} before the transformation.
Samples prepared at constant pressure follow again the trends of the hydrostatically compressed glasses but without the abrupt change in behavior.

\subsection{Elastic instabilities and soft spots}
Although the (ensemble-averaged) shear modulus softens under compression, its value at the critical pressure for the phase transformation is still finite.
The material does not appear to exhibit an elastic instability.
While the ensemble-averaged data does not show an instability, the individual calculations undergo a continuous series of elastic instabilities, i.e. a divergence of the shear modulus in the pressure range of the polyamorphous transition. 
Figure~\ref{fig: KG_coefficients}c shows the non-affine displacement field resulting from one of these elastic instabilities.
This non-affine displacement field reveals a random displacement of all atoms, but localized on one cluster of atoms. 
Figure~\ref{fig: KG_coefficients}d shows the drops in the shear modulus resulting from a sequence of such events.
Comparing the shear modulus before and after an instability, we observe that the absolute value barely changes for pressures $P < P^{\text{c}}$.
At the critical pressure $P^{\text{c}}$ for the transition (which is slightly configuration dependent), the number of elastic instabilities per pressure range increases.
Interestingly, each instability near $P^\text{c}$ lead to a noticeable stiffening of the shear modulus.

We investigate the origin of this sudden change by looking at the microscopic signature of these individual plastic events, which are reminiscent of shear transformation zones (STZs) but occur under compression rather than shear.
Many methods exist to predict and identify soft spots in amorphous solids~\cite{richard_plasticity_2020}.
In this work we use the method developed by Richard et al.~\cite{richard_STZ_2021},
which relies on the pseudo harmonic modes (PHMs).
We choose PHMs because they predict the displacement field of a soft spot well before the instability~\cite{richard_STZ_2021}.  
We compute the lowest PHM for a subset of 50 configurations as a function of hydrostatic pressure, and report their approximate size using the participation ratio.
The participation ratio $e$ is defined as
\begin{equation}
    N e
    =
    \left[
    \sum_i
    \left(
    \vv{\pi}_i \cdot \vv{\pi}_i
    \right)^2
    \right]^{-1},
\end{equation}
where $\boldsymbol{\pi}$ (which is a $3N$-dimensional vector) is the displacement field of the PHM and $N$ is the number of atoms in the solid.
The participation ratio $e$ is a measure of spatial localization and $Ne$ is the approximate number of atoms participating in the plastic event.
For the special case where only one atom is involved in a plastic event, it yields $Ne = 1$.
In contrast, if all atoms are involved and they contribute equally, $Ne = N$.

\begin{figure}[t!]
    \includegraphics[]{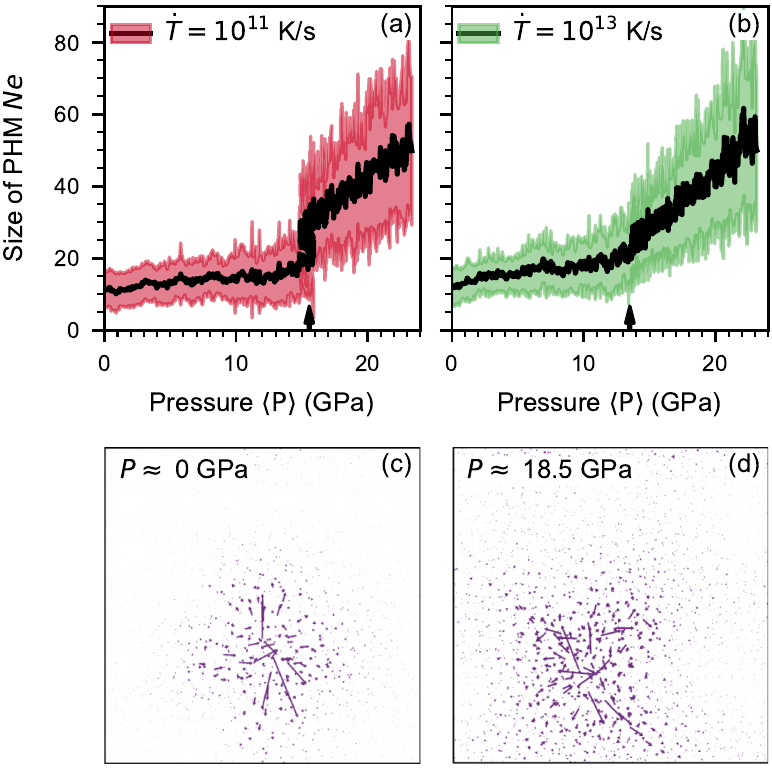}
    \caption{
    \label{fig:size_stz} 
    (a),(b) Pressure dependence of the number of atoms contributing in one pseudoharmonic mode (PHM).
    The black vertical arrows mark the critical pressures $P^\text{c}$ from Fig.~\ref{fig: hydrostatic_compression}b.
    (c),(d) Displacement fields $\vv{\pi}_i$ for PHMs for one sample prepared at $\dot{T}=10^{13}\,$K/s.
}
\end{figure}

Figures~\ref{fig:size_stz}a and b show the product $Ne$ of the PHM as a function of hydrostatic pressure.
The mean size of the PHMs increases slightly with compression before the phase transformation.
Independent of quench rate, the mean size of the PHMs in this regime involves approximately $10-20$ atoms.
At the transition pressure, the mean size of the soft spots starts to increase up to $\approx 30\,$atoms.
For larger compression, the number of atoms involved in a plastic event increases up to approximately $50$ atoms.
The displacement field $\boldsymbol{\pi}$ for one of the modes in the low and high pressure regime is shown in Fig.~\ref{fig:size_stz}c and d.
It is clearly visible that the size of the soft spots increases with compression and that the displacement field has an Eshelby-like character~\cite{eshelby_1957,albaret_mapping_STZ_2016,richard_plasticity_2020}. 

\section{Discussion}

First, we note that equilibrium arguments appear to be incapable of capturing the polyamorphic transition in aSi.
In particular, the polyamorphic transition does not occur near the ``equilibrium'' pressure $P^\text{eq}$.
However, $P^\text{eq}$ does capture aspects of the transition from low- to high-density structures when compressing the melt before quenching -- in particular the maximum in shear modulus (Fig.~\ref{fig: KG_coefficients}b).
This indicates a large kinetic barrier resisting the structural transition, which emphasizes its nonequilibrium character.

Our observations on the polyamorphic transition have similarities to shear yielding of glasses.
Shear-yielding of aSi has been reported in the literature, with the a phenomenology as follows: aSi initially deforms elastically but flows after the yield point~\cite{demkowicz_plastic_carries2_2005,fusco_shear_aSi_2010,kerrache_aSi_2011,albaret_mapping_STZ_2016,boioli_STZ_2017};
approaching the yield point, a large percentage of atoms become $5$-fold coordinated~\cite{demkowicz_plastic_carries2_2005,fusco_shear_aSi_2010,kerrache_aSi_2011};
the shear modulus changes abruptly at the yield point~\cite{albaret_mapping_STZ_2016}.
We observed the same phenomena for the polyamorphic transition, with the difference that atoms with coordination larger than 5 appear during compression.

The microscopic picture of the polyamorphic transition found here is also identical to yielding under shear in glasses.
Hydrostatic compression of aSi results in series of instabilities, which can be identified by loss of mechanical stability that manifests as a vanishing shear modulus.
Each instability triggers a ``soft'' spot in the material:
The displacement field during these instabilities shows signatures identical to STZs~\cite{tanguy_continuum_limit_2002,maloney_breakdown_2004,richard_plasticity_2020}, the carrier of plastic deformation in disordered materials under shear.
These signatures include a localization of the non-affine displacement field~\cite{falk_dynamics_1998} as well as a quadrupolar strain-field surrounding the localized event, as described by Eshelby inclusion theory~\cite{eshelby_1957}.
The fast-quenched configurations initially have a larger fraction of $5$-fold and higher-coordinated atoms than the configurations obtained at lower quench rates.
These are ``liquid-like'', i.e. less resistant to deformation and acting as soft spots in aSi~\cite{demkowicz_plastic_carriers1_2004,demkowicz_plastic_carries2_2005}.
This lower resistance to deformation reduces the critical pressure.

The macroscopic shear modulus, computed here as ensemble averages over $500$ configurations, does not vanish.
However, it monotonously decreases with pressure below $P^\text{c}$ of the polyamorphic transition and jumps to a larger value as the transition is crossed.
This is accompanied by an instantaneous change in density, marked by an increase in average coordination.
The decrease in shear modulus is driven by the non-affine contribution to the elastic modulus (Fig.~\ref{fig:contribution_Kmu}e).
In terms of the microscopic picture painted in the previous paragraph, the non-affine origin of the decrease in shear modulus can be interpreted as a signature of localized softening of the material.

\begin{figure}[t!]
    \includegraphics[]{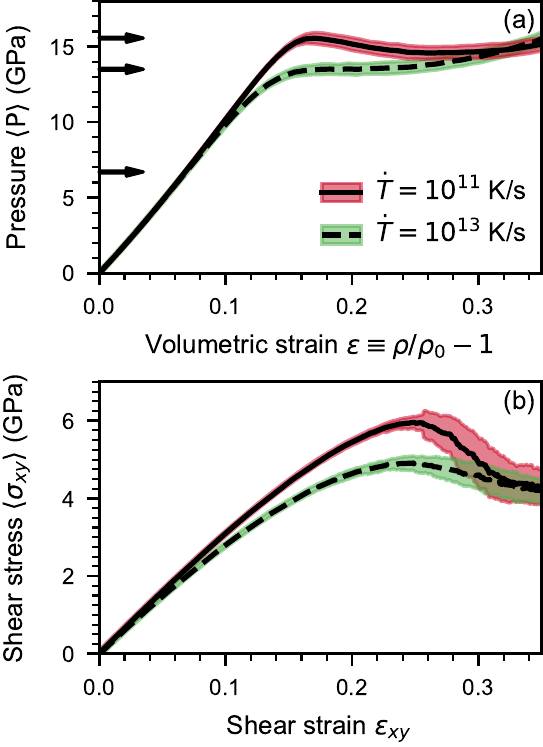}
    \caption{
     (a) Hydrostatic pressure as a function of volumetric strain $\epsilon$.
     The volumetric strain is computed as $\epsilon = \rho / \rho_0 - 1$, where $\rho_0$ is the initial density.
     (b) Shear stress as a function of shear strain.
     }
    \label{fig: strain_pressure}
\end{figure}

To further emphasize the similarities to yielding, we plot in Fig.~\ref{fig: strain_pressure}a the pressure as a function of volumetric strain in a traditional stress-strain diagram.
The behavior is qualitatively identical to stress-strain diagrams obtained from shear yielding glasses~\cite{maloney_athermal_2006,demkowicz_plastic_carriers1_2004,demkowicz_plastic_carries2_2005,fusco_shear_aSi_2010,kerrache_aSi_2011,Ozawa2018-fw}, with pressure taking the role of shear stress.
For visual comparison, we also computed the shear response of our aSi samples and show in Fig.~\ref{fig: strain_pressure}b the ensemble-averaged shear stress $\sigma_{xy}$ as a function of shear strain $\varepsilon_{xy}$.
The material exhibits a linear ``elastic'' response at small strain $\varepsilon$ followed by flow at constant pressure beyond a yield point at $\varepsilon\gtrsim 0.1$ und both compression and shear.
In particular, the glass prepared at a lower quench rates shows a stress overshoot, explaining the reentrant section of the density-pressure diagram (Fig.~\ref{fig: hydrostatic_compression}b) as a compression-softening phenomenon of a well-aged glass~\cite{varnik_static_yield_2004,demkowicz_plastic_carries2_2005,Ozawa2018-fw}.

\section{Conclusions}
In summary, we performed extensive simulations of the behavior of aSi during hydrostatic compression.
Our simulations show a polyamorphic transition from a low-density to a high-density phase during quasistatic compression of aSi quenched at zero pressure, but a gradual transition for near-equilibrium structures obtained when quenching liquid silicon at finite pressure.
Microscopic and macroscopic analysis of the transition revealed similarities to the yield transition of glasses under shear: Localized carriers of plastic events (shear transformation zones), change of local atomic order, and change in elastic properties.
We conclude that the polyamorphic transition is essentially a yield transition.
Whether yield of glasses itself is a phase transition is an open discussion in the scientific literature~\cite{wisitsorasak_RFOT_Yield_2012,jaiswal_yield_first_order_2016,Parisi2017-gu,Ozawa2018-fw,Jana2019-eb}.

\section*{Acknowledgment}
We thank Richard Leute and Thomas Reichenbach for fruitful discussions.
Molecular dynamics simulations were carried out with \textsc{lammps}~\cite{plimpton_fast_algorithms_1995, thompson_lammps_2022}. 
We used \textsc{ase}~\cite{hjorth_larsen_atomic_2017} and \textsc{matscipy} for analysis of results and computation of the elastic properties.
Atomic configurations were visualized with \textsc{ovito}~\cite{stukowski_ovito_2009}.
The authors acknowledge support from the Deutsche Forschungsgemeinschaft (DFG, grant PA 2023/2).
Simulations were carried out on NEMO at the University of Freiburg (DFG grant INST 39/963-1 FUGG).


\providecommand{\noopsort}[1]{}\providecommand{\singleletter}[1]{#1}%

\end{document}